\documentclass[a4paper,UKenglish]{lipics}

\usepackage{microtype}

\usepackage[T1]{fontenc}
\usepackage{amsmath}
\usepackage{amssymb}
\usepackage{url}
\usepackage{framed}
\usepackage{hyperref}
\usepackage[bottom]{footmisc}

\newcommand{\imp}{\ensuremath{\rightarrow}}

\newcommand{\yields}{\ensuremath{\Longrightarrow}}

\newcounter{TODOcounter}

\begin{document}
 
%\title{Teaching Sequent Calculus: Advantages of Using an Interactive Trainer}
\title{The Sequent Calculus Trainer -- Helping Students to Correctly Construct Proofs}

\author{Arno Ehle}
\author{Norbert Hundeshagen}
\author{Martin Lange}
%\affil{Fachbereich Elektrotechnik/Informatik, \\ Universit\"at Kassel, Germany \\
%       \texttt{\{ehle,hundeshagen,lange\}@uni-kassel.de} }
\affil{School of Electrical Engineering and Computer Science \\ University of Kassel, Germany 
%\\   %\texttt{\{ehle,hundeshagen,mlange\}@uni-kassel.de} 
}
\maketitle

\Copyright{Arno Ehle and Norbert Hundeshagen and Martin Lange}%mandatory, please use full first names. LIPIcs license is "CC-BY";  http://creativecommons.org/licenses/by/3.0/

\serieslogo{logo_ttl}%please provide filename (without suffix)
\volumeinfo%(easychair interface)
  {M. Antonia {Huertas}, Jo\~ao {Marcos}, Mar\'ia {Manzano}, Sophie {Pinchinat}, \\
  Fran\c{c}ois {Schwarzentruber}}% editors
  {5}% number of editors: 1, 2, ....
  {4th International Conference on Tools for Teaching Logic}% event
  {1}% volume
  {1}% issue
  {35}% starting page number
\EventShortName{TTL2015}
%\DOI{10.4230/LIPIcs.xxx.yyy.p}% to be completed by the volume editor

\begin{abstract}
We present the Sequent Calculus Trainer, a tool that supports students in learning how to correctly
construct proofs in the sequent calculus for first-order logic with equality. It is a proof assistant fostering
the understanding of all the syntactic principles that need to be obeyed in constructing correct proofs. It 
does not provide any help in finding good proof strategies. Instead it aims at understanding the sequent calculus
on a lower syntactic level that needs to be mastered before one can consider proof strategies. Its main feature is a proper
feedback system embedded in a graphical user interface.

We also report on some empirical findings that indicate how the Sequent Calculus Trainer can improve the students'
success in learning sequent calculus for full first-order logic.     
\end{abstract}

\section{Introduction}
Many courses in theoretical computer science suffer from high failure rates and it is common among students to alienate themselves from such courses,
c.f.\ \cite{Robins88,Surma:2012:RDT:2379703.2379716}. 
The reasons are manifold and vary from the way mathematics is taught in school, to lack of general problem solving competences, 
and not least, to a diversity in skills for adapting new knowledge.

The BSc computer science curriculum at the University of Kassel contains a 2nd-year mandatory course on logic in computer science, as it can be found
in typical computer science university programs. This course has been re-organised in recent years with the aim of improving learning
outcomes and therefore reducing failure rates. A central point of this re-organisation is the use of constructivistic learning theory, in particular 
the inverted-classroom model (c.f.\ \cite{LPT2000}). This model focuses on learning, literally, as a self-organised activity; consequently, the logic 
course should engage students with methods and tools to assist and self-assess the use of formal logic and the calculi taught with them. One of these 
tools, developed for such purposes, is the \emph{Sequent Calculus Trainer} whose design and use will be described in this paper.

We start by explaining typical problems that students encounter when being faced with a standard exercise in learning Gentzen's sequent calculus
\cite{Gentzen35}: to find a proof for a given sequent, formally expressing that some formula is a logical consequence of a set of formulas. We assume 
the reader to be entirely familiar with first-order logic with equality \cite{eft_logic_book84} and proof calculi in general. Familiarity with 
sequent calculus in particular is not strictly necessary to follow those explanations; the principles should become clear from the examples we
use. We give a brief description of the Sequent Calculus Trainer and explain its aims. At last we provide some empirical data
that supports the claim that the Sequent Calculus Trainer can effectively aid the learning of the sequent calculus for first-order logic.

\section{The Sequent Calculus}

\subsection{The Proof Rules}
The sequent calculus is a system of proof rules that operate on \emph{sequents} which are pairs of multi-sets of formulas, written 
$\Gamma \Longrightarrow \Delta$. The intended meaning of such a sequent is that the conjunction over $\Gamma$ logically implies the disjunction
over $\Delta$. The rules listed in Figure~\ref{fig:rules} operate on formulas in the antecedent $\Gamma$ or succedent $\Delta$ of the rule's 
conclusion below the line, possibly producing new premisses shown above the line. A proof for a sequent $\Gamma \Longrightarrow \Delta$ is, as 
usual, a finite tree of sequents formed from $\Gamma \Longrightarrow \Delta$ at its root by successively applying these rules. Each branch must end in an
instance of an axiom, i.e.\ a rule with no premisses.

In the rules for quantified formulas, $c$ must always be a fresh constant -- called Skolem constant -- that does not occur in the conclusion of 
this rule already; $t$ must be a ground term over the symbols that occur in the conclusion.

\begin{figure}[t]
\begin{displaymath}
% And left
\begin{array}{c}
\Gamma, \varphi, \psi \Longrightarrow \Delta \\ \hline
\Gamma, \varphi \wedge \psi \Longrightarrow \Delta
\end{array}
\qquad
%
% And right
\begin{array}{c}
\Gamma \Longrightarrow \varphi, \Delta \qquad \Gamma \Longrightarrow \psi, \Delta \\ \hline
\Gamma \Longrightarrow \varphi \wedge \psi, \Delta
\end{array}
\qquad
%
% Neg right
\begin{array}{c}
\Gamma, \varphi \Longrightarrow \Delta \\ \hline
\Gamma \Longrightarrow \neg\varphi, \Delta
\end{array}
\qquad
%
% Neg left
\begin{array}{c}
\Gamma \Longrightarrow \varphi, \Delta \\ \hline
\Gamma, \neg\varphi \Longrightarrow \Delta
\end{array}
\end{displaymath}
\begin{displaymath}
% Or left
\begin{array}{c}
\Gamma, \varphi \Longrightarrow \Delta \qquad \Gamma, \psi \Longrightarrow \Delta \\ \hline
\Gamma, \varphi \vee \psi \Longrightarrow \Delta
\end{array}
\qquad
%
% Or right
\begin{array}{c}
\Gamma \Longrightarrow \varphi, \psi, \Delta \\ \hline
\Gamma \Longrightarrow \varphi \vee \psi, \Delta
\end{array}
\qquad
%
% Contr right
\begin{array}{c}
\Gamma \Longrightarrow \varphi,\varphi,\Delta \\ \hline
\Gamma \Longrightarrow \varphi, \Delta
\end{array}
\qquad
%
% Contr left
\begin{array}{c}
\Gamma, \varphi, \varphi \Longrightarrow \Delta \\ \hline
\Gamma, \varphi \Longrightarrow \Delta
\end{array}
\end{displaymath}
\begin{displaymath}
% Imp left
\begin{array}{c}
\Gamma, \psi \Longrightarrow \Delta \qquad \Gamma \Longrightarrow \varphi, \Delta \\ \hline
\Gamma, \varphi \to \psi \Longrightarrow \Delta
\end{array}
\quad\enspace
%
% Imp right
\begin{array}{c}
\Gamma, \varphi \Longrightarrow \psi, \Delta \\ \hline
\Gamma \Longrightarrow \varphi \to \psi, \Delta
\end{array}
\quad\enspace
%
% Ax
\begin{array}{c}
\qquad \\ \hline
\Gamma, \varphi \Longrightarrow \varphi, \Delta
\end{array}
\quad\enspace
%
% Eq right
\begin{array}{c}
\qquad \\ \hline
\Gamma \Longrightarrow s=s, \Delta
\end{array}
\end{displaymath}
\begin{displaymath}
% Ex left
\begin{array}{c}
\Gamma, \varphi[c/x] \Longrightarrow \Delta \\ \hline
\Gamma, \exists x\;\varphi \Longrightarrow \Delta
\end{array}
\qquad
%
% Ex right
\begin{array}{c}
\Gamma \Longrightarrow \varphi[t/x], \Delta \\ \hline
\Gamma \Longrightarrow \exists x\;\varphi, \Delta
\end{array}
\qquad
%
% All left
\begin{array}{c}
\Gamma, \varphi[t/x] \Longrightarrow \Delta \\ \hline
\Gamma, \forall x\;\varphi \Longrightarrow \Delta
\end{array}
\qquad
%
% Ex right
\begin{array}{c}
\Gamma \Longrightarrow \varphi[c/x], \Delta \\ \hline
\Gamma \Longrightarrow \exists x\;\varphi, \Delta
\end{array} 
\end{displaymath}
\begin{displaymath}
% Eq left
\begin{array}{c}
\Gamma, s=s \Longrightarrow \Delta \\ \hline
\Gamma \Longrightarrow \Delta
\end{array}
\qquad
%
% Subst left
\begin{array}{c}
\Gamma, \varphi[s'/x] \Longrightarrow \Delta \\ \hline
\Gamma, s=s', \varphi[s/x] \Longrightarrow \Delta
\end{array}
\qquad
%
% Subst right
\begin{array}{c}
\Gamma \Longrightarrow \varphi[s'/x], \Delta \\ \hline
\Gamma, s=s' \Longrightarrow \varphi[s/x], \Delta
\end{array}
\end{displaymath}
\caption{The proof rules of the sequent calculus.}
\label{fig:rules} 
\end{figure}

\subsection{A Didactic Perspective}
\label{didactics}

A standard exercise in sequent calculus asks for a proof of a given sequent, e.g.\ $\mathcal{S}_0 :=$
\begin{displaymath}
\forall x\forall y.\; E(x,y) \imp x = f(y) ~ \yields ~ \forall x\forall y\forall z.\; E(x,z) \wedge E(y,z) \imp x = y\ .
\end{displaymath}
%Starting with this sequent, one needs to apply rules and axioms -- there are less than twenty -- to build a finite proof tree for this sequent.
Difficulties and mistakes can generally be put into two categories. 
\medskip

(1) The first one is about \emph{constructing a correct proof}: many students are not able to handle formalisms well; often they can barely parse  
sequents and apply rules correctly. In this example, one has to introduce new names for the universally quantified variables $x,y,z$ in this order and 
then decompose the Boolean operators on the right side, yielding $\mathcal{S}_1 :=$
\begin{displaymath}
\forall x\forall y.\; E(x,y) \imp x = f(y), E(a,c), E(b,c) ~ \yields ~  a = b\ .
\end{displaymath}
Typical mistakes at this syntactic level are concerned with wrong rule applications and include 
\begin{itemize}
\item \emph{confusing} rules, for instance applying the rule for conjunctions to a disjunction;
\item \emph{misplacing} rules, usually by applying a rule to a genuine subformula rather than a formula in the sequent; in other words not
      understanding the structure of a sequent;
\item \emph{wrong first-order instantiations}, for instance not choosing a fresh Skolem constant when needed;
\item \emph{wrong rule instantiations}, for instance by adding the symbols $\Gamma$ and $\Delta$ to the sequent at hand; 
\end{itemize}
and so on.
\medskip

(2) There are other ways of applying rules in a syntactically correct way in this example, for instance by operating on the formulas in the premiss 
instead. This, however, is unwise for \emph{finding the right proof}. For this one must proceed as described above and then have the inspiration to 
double one of the formulas in the premiss and instantiate both copies differently yielding
\begin{displaymath}
E(a,c) \imp a = f(c), E(b,c) \imp b = f(c), E(a,c), E(b,c) ~ \yields ~  a = b\ .
\end{displaymath}
The rest of this proof task is relatively easy using Boolean elimination rules and very simple reasoning about equalities. 
\medskip

There is a clear dependency between these two challenges: the ones described in (1) need to be met before those in (2) can be met; it is impossible to
find a proof unless one is able to construct correct proofs at all. The latter is clearly a very difficult task for students who already struggle with 
uncertainties like ``am I allowed to apply this rule here?'', ``was the application correct?'', ``should I introduce a new name or instantiate with 
an already existing term?'', etc.

This gap between syntactical and semantical understanding has been addressed in many
fields like teaching programming or simply mathematics (e.g.\ in \cite{DBLP:journals/ijpp/ShneidermanM79}).  This phenomenon is accurately described in
\cite{DBLP:conf/ticttl/GasquetSS11a} as the ability to ``write rather rigorously a simple C program'', while they cannot ``rigorously write down a
mathematical proof of the kind needed in graph theory, formal logic, [...]''  There is a hidden hint in this observation on how to tackle the problem
of teaching proof calculi. Students seem to easily understand syntactic principles as long as there is a mechanism -- like a compiler -- which allows them 
to learn the formalism in a trial-and-error way.

This is where the Sequent Calculus Trainer comes into play. It is supposed to aid the constructing of correct proofs but does not help at all in finding 
the right proofs. From a logical point of view it is merely a proof assistant, not a theorem prover. From a didactic point of view, its use is supposed
to address the students' rote memory such that they become able to reach this gap between syntax and semantics, thus enabling them to start understanding
the underlying logical concepts. The Sequent Calculus Trainer therefore adheres to two design principles: it is an easy-to-use and simple assistant for
building proof trees in sequent calculus. Moreover, it comes with a compiler-like feedback system, which is known for its benefits in tutorial
teaching environments \cite{Anderson1995}.
 
\subsection{Related Tools}
There are other high quality interactive proof systems that can be used to train the construction of correct proofs in the sequent calculus. The tool that
meets the prerequisites laid out here best is LOGITEXT\footnote{\url{http://logitext.mit.edu}}; others worth mentioning are JAPE \cite{DBLP:journals/cj/BornatS99} and PANDA
\cite{DBLP:conf/ticttl/GasquetSS11a}. None of these treats first-order logic \emph{with} equality, though. This is a major drawback since equality is --
possibly together with quantification -- one of the most difficult concepts for constructing and finding proofs. Yet equality reasoning is ubiquitous in
computer science; hence, it should not be ignored in teaching contexts. Furthermore, JAPE is lacking a feedback system, and PANDA's proof calculus is
natural deduction which differs from the sequent calculus. 

In a setting where we measure success through the understanding of syntactic concepts, it is essential that the proof calculus used in classroom must
be the same as that used by a supporting tool. None of those existing tools is general enough to serve the purposes described above -- to act as a tool
supporting the learning of proof construction in the sequent calculus for first-order logic with equality. One also should not underestimate the effort 
that would be needed in order to extend or amend an existing software tool created by others. Hence, having many tools with slightly differing features
in this area should be considered advantageous.

%Given the students' aforementioned problems on a purely syntactic level already, 
\begin{figure}
\begin{center}
\includegraphics[width=\textwidth]{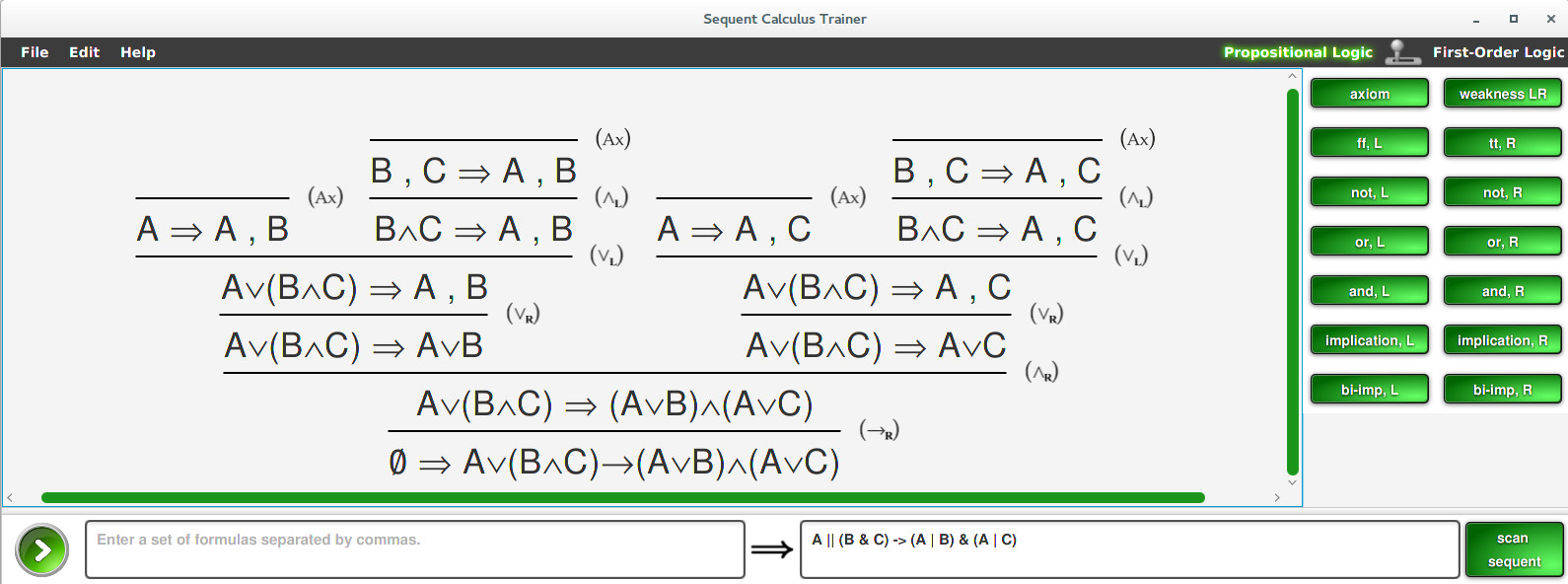}
\caption{The user-interface.}\label{ui}
\end{center}
\end{figure}

\section{The Sequent Calculus Trainer}
\begin{figure}[!t]
\begin{center}
\includegraphics[width=\textwidth]{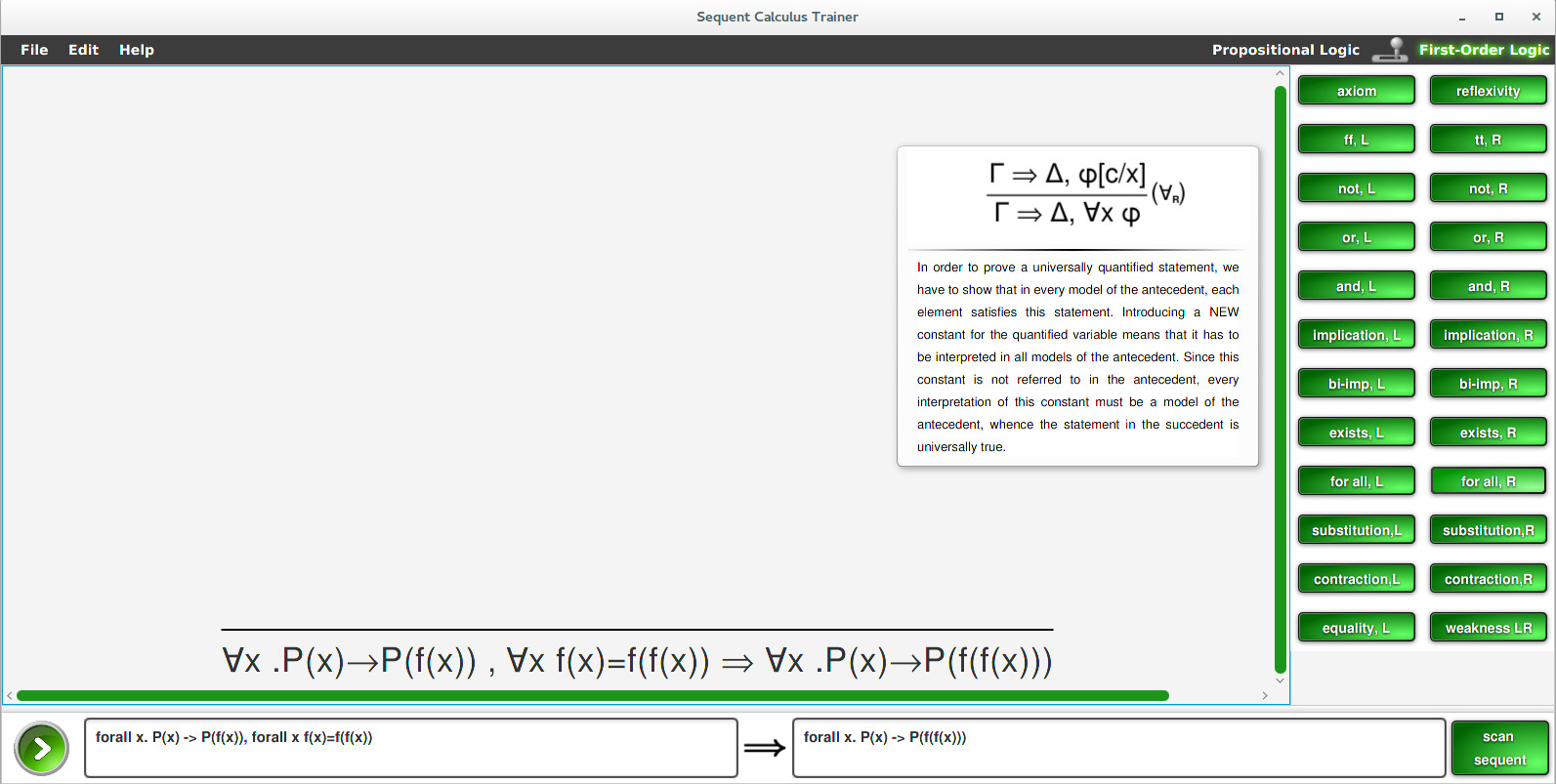}

\vspace*{0.8em}
\includegraphics[width=\textwidth]{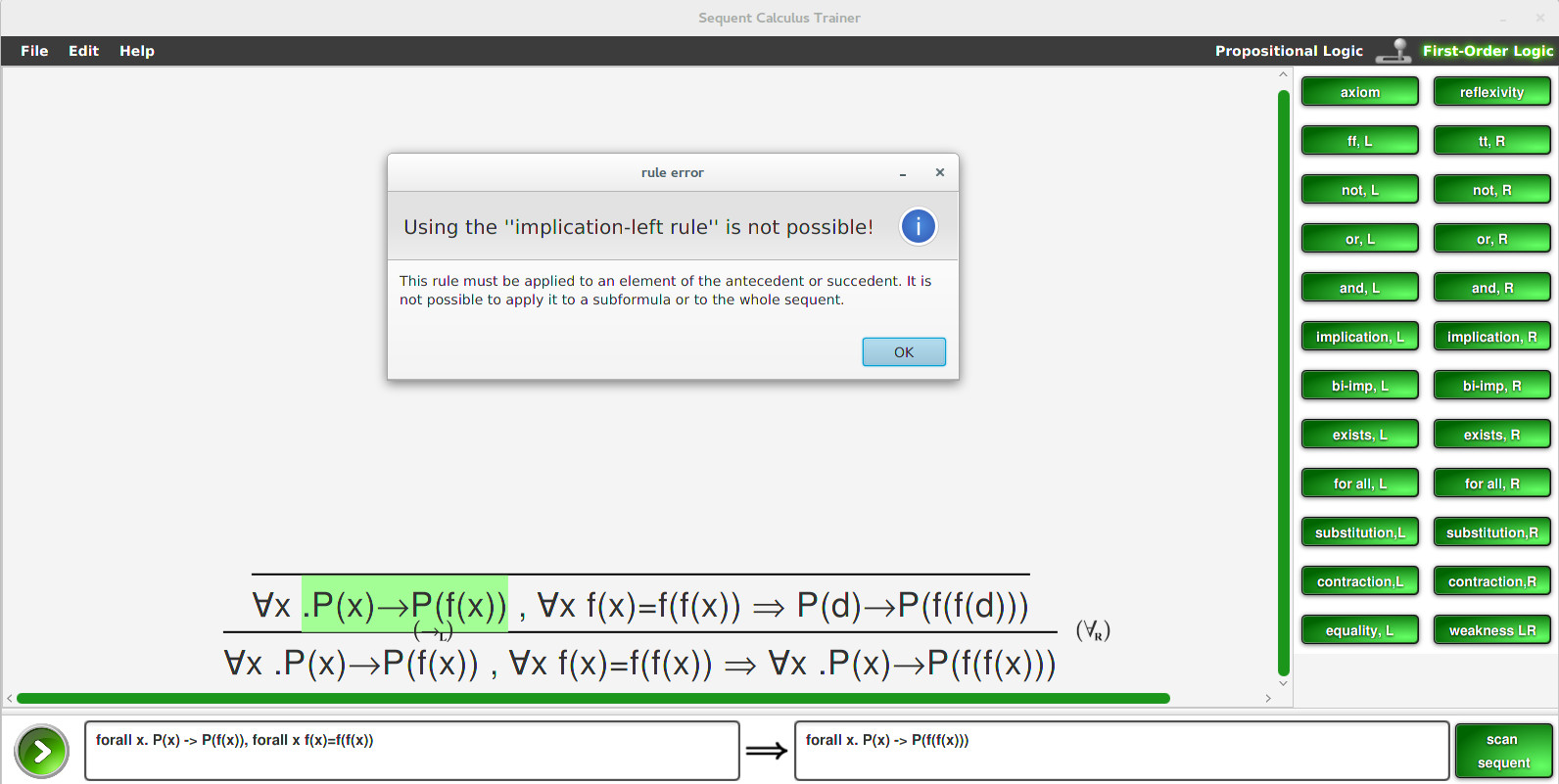}
\caption{The feedback system.}\label{feedback}
\end{center}
\end{figure}
We provide the Sequent Calculus Trainer as an open source application under the BSD-3 license and the source code as well as the binaries are publicly
available\footnote{\url{http://www.uni-kassel.de/eecs/fachgebiete/fmv/projects/sequent-calculus-trainer.html}}. Figure~\ref{ui} shows the graphical user interface which is kept fairly
simple. The trainer is shipped with two main views, one for propositional logic and one for first-order logic. They both differ only in the number of
applicable rules shown on the right side of the windows, and in the treatment of atomic propositions, which are interpreted as 0-ary predicates in the first-order logic view. Sequents can be input either through a simple text file or in the text fields at the bottom,
where the syntax specification for the input is given, too. Furthermore, it is possible to save and load proof trees in an internal format as well as
export them in PNG format.

\subsection{Key Features}

\begin{figure}[!t]
\begin{center}
\includegraphics[width=\textwidth]{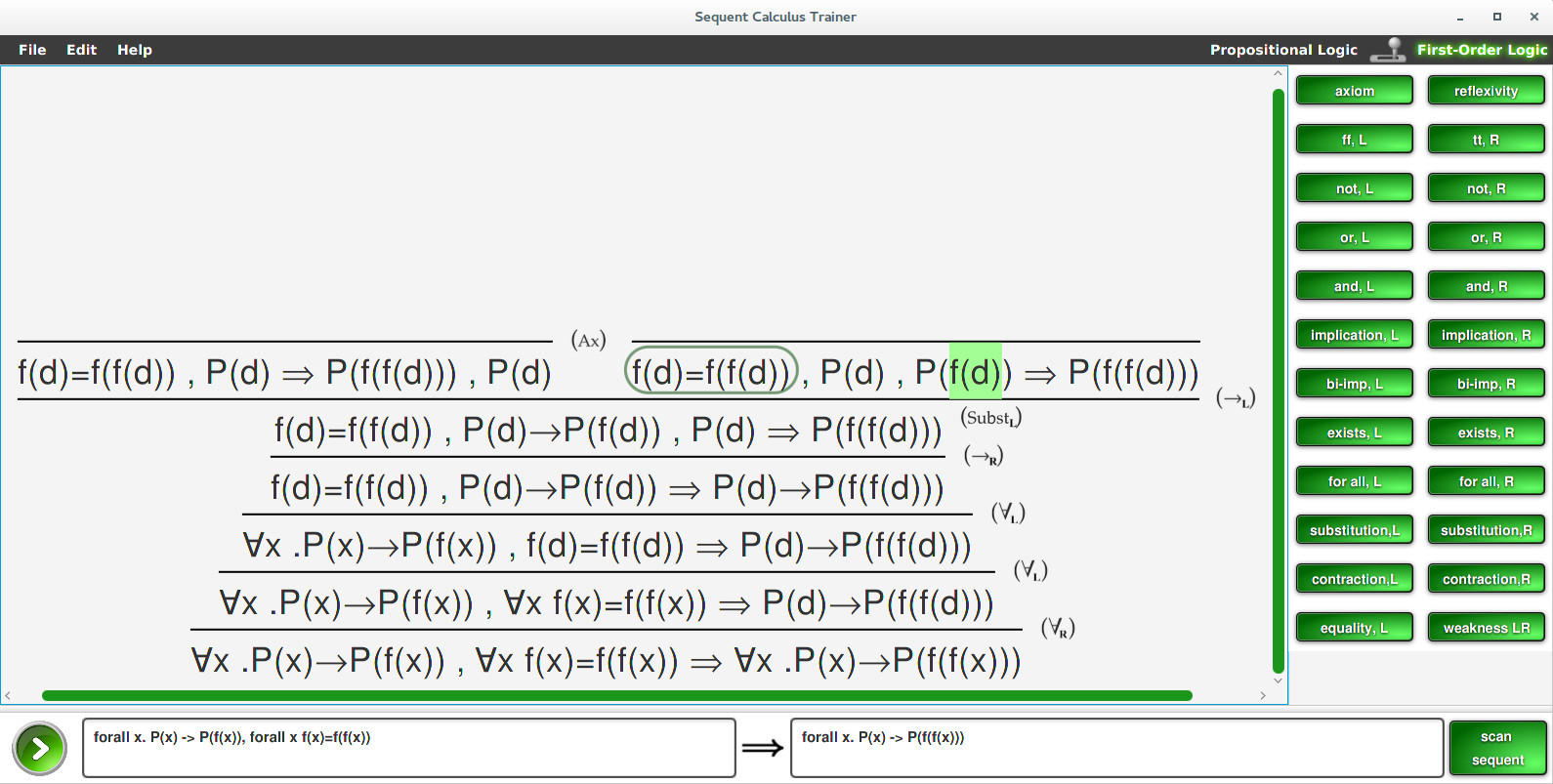}

\vspace*{0.8em} 
\includegraphics[width=\textwidth]{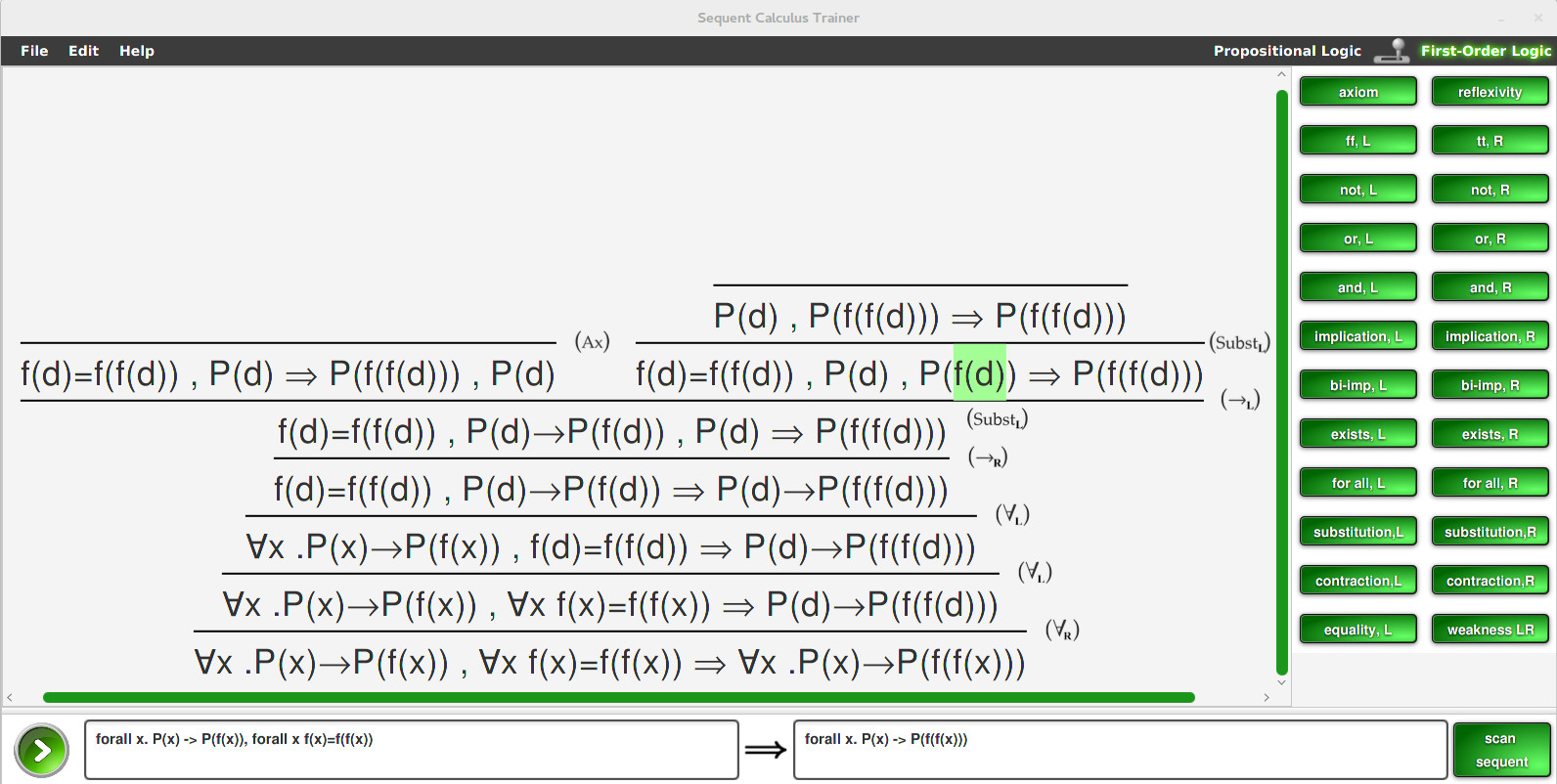}
\caption{The substitution rule.}\label{equality}
\end{center}
\end{figure}
We briefly introduce the two key features of the Sequent Calculus Trainer. 
\begin{itemize}
\item Nearly every user action leads to a response by the program. Figure~\ref{feedback} exemplarily illustrates such on-screen messages. Each rule
  button is equipped with a short message, which occurs on mouse-over. These messages usually contain the formal definition of a rule as well as
  some appropriate high-level explanations of the rule's meaning and, if suitable, why it is a valid logical principle. 

  If the user has chosen a rule and tries to apply it to a formula by selecting a logical operator, the formula represented by this operator ``responds'' 
  by telling the user whether or not the rule is applicable there. This happens in two ways: the part of the formula that is in scope of the selected 
  operator or symbol is highlighted. This helps to understand precedence rules and the structure of sequents and formulas.  When a wrong operator or 
  symbol is chosen, the user is provided with an error message which includes a hint on the mistake. For instance, the reason why a current leaf in 
  the prooftree is an axiom has to be identified via clicking on the part of the formula that causes the application of an axiom rule.

\item The second notable feature is the handling of sequents that include equalities. Figure~\ref{equality} shows how the substitution rule works. 
  After the rule for substitution on the left-hand or right-hand side of the sequent has been chosen, the program expects an atomic formula with an 
  equality predicate to be selected. In the last step, the term that should be substituted needs to be clicked on.
\end{itemize}
A final point worth mentioning is that the user is able to undo all steps in the proof up to a certain sequent at any time by just applying a different 
rule to that particular sequent.

\subsection{The Backend} 

The Sequent Calculus Trainer is meant to be used by a broad range of students; hence, it platform independency provided by an implementation in Java is a
key design principle. It uses the GUI framework JavaFX, which is integrated in the Java Standard Library since the emerging of Oracle's Java 8. This mainly
allows the program to have only one dependency in ControlsFX that is a small extension library for JavaFX, designed to give even more UI controls and
simple dialogs. The result is a clear and comprehensive feedback system which was another key design principle.

The Sequent Calculus Trainer is designed according to the principles provided in the model-view-controller-patterns (MVC). Thus, it consists of three
parts; the data model which contains the data and algorithms as a standalone part of the program, the controllers which contain the logic for the GUI,
and finally the GUI itself which is build with JavaFX while mainly using the XML based notation for GUI elements called FXML. 
The program is equipped with the possibility to simply add new languages in form of language sets in simple text files and new country flags for the 
frontend. Currently German and English are available.

\section{Experiences in Teaching the Sequent Calculus}
\label{sec:empirical}

The computer science BSc curriculum at the University of Kassel contains a mandatory 2nd-year course on logic which teaches, amongst others, the sequent
calculus for first-order logic with equality. In order to pass the course students need to take a written exam. We report on the results achieved in
three successive years. Group 1 took the exam in winter term 2012/13. The Sequent Calculus Trainer was developed afterwards, so it was not available to
them at the time of preparation for the exam. Group 2 took the exam in winter term 2013/14, and they were greatly encouraged to solve corresponding
exercises using the Sequent Calculus Trainer. Group 3 took the exam in winter term 2014/15. The Sequent Calculus Trainer was made available, but in
comparison to group 2 the trainer had been advertised less. The preconditions are comparable in the sense that before the exam, all groups had to
pass two graded exercises on sequent calculus, they were provided with the same number of additional voluntary tasks on this topic, and they were allowed
to prepare a handwritten note for the exam which contained the sequent calculus rules in many cases.

\begin{figure}
\begin{center}
%\includegraphics[width=\textwidth]{example_solutions/exsol1.jpg}
%\vspace*{0.1em} 
\includegraphics[width=\textwidth]{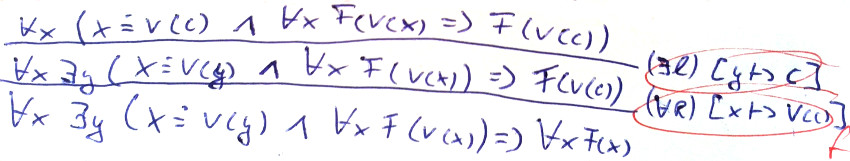}
\vspace*{.5em}
\rule{\textwidth}{.5pt}

\vspace*{.5em}
\includegraphics[width=\textwidth]{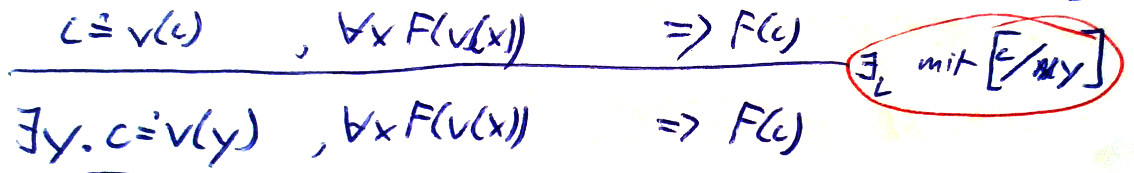}
\vspace*{.5em}
\rule{\textwidth}{.5pt}

\vspace*{.5em}
\includegraphics[width=\textwidth]{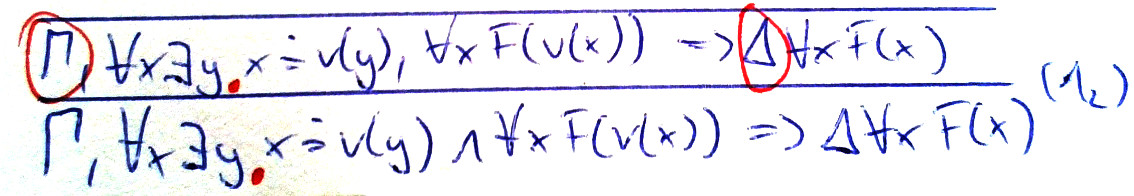}
\end{center}
\caption{Examples of frequently made mistakes with wrong rule applications.}
\label{fig:solutions}
\end{figure}

The three exams under consideration featured a question each, asking for a proof of a given sequent. The three sequents differed, 
with the ones used for group 2 and 3 being seemingly more difficult to prove with regards to both aspects of constructing and finding a proof. 
\begin{center}
\begin{tabular}{ll}
group 1: & $\forall x\exists y.\; x = v(y) \wedge \forall x\; F(v(x)) \enspace \Longrightarrow \enspace \forall x\; F(x)$ \\
group 2: & $\forall x\forall y.\; E(x,y) \imp x = f(y) ~ \yields ~ \forall x\forall y\forall z.\; E(x,z) \wedge E(y,z) \imp x = y$ \\
group 3: & $\forall x\forall z.\; P(x,c) \wedge Q(z,g(x,z)) ~ \yields ~ \exists y\forall x\; P(x,y) \wedge \forall z\exists u\; Q(z,u)$ 
\end{tabular}
\end{center}
While the sequents for groups 2 and 3 need some insight into the properties expressed by its formulas, the sequent for group 1 can be proved using almost
syntactic considerations only. Surprisingly, out of 70 students in group 1, more than 50\% (46 in total) were not able to apply such a simple strategy of
applying all possible rules in the correct order. Moreover, when adding the 10 students that did not even try to execute the exercise, we may conclude
that 80\% of group 1 did not succeed in this exercise because of problems with the correct application of rules. The mistakes most frequently made
were \emph{misplacing} of rules and \emph{wrong first-order instantiations} according to the classification listed in Section~\ref{didactics}. Some examples of typical mistakes are shown in Figure~\ref{fig:solutions}.

Group 2 shows a totally different picture. 
%, which was provided with the sequent calculus trainer. 
Out of 51 students only 12 already failed in simple rule applications, where 1 student did not execute the exercise, which leads to 25\% in total. Thus,
nearly 75\% of the students where at least able to handle the syntactic formalism resulting in correct rule applications before a deeper 
understanding would be needed. 

The most recent data is taken from group 3. Out of 46 students 19 failed to achieve the mentioned goal and 2 did not execute the exercise which is 46\%
in total.

If we summarize the results of the latter two groups we get a rate of 34 students out of 97 which failed for reasons of wrong application of rules. Thus,
nearly 65\% of the students where at least able to handle the syntactic formalism resulting in correct rule applications. We believe that this is due to 
the availability of the Sequent Calculus Trainer.

This conclusion seems to be in contrast to two other -- seemingly odd -- observations. First of all, group 2 did not achieve significantly more points than group 1.
Both exercises were graded with 4 points in total where group 1 reached an average of 1.2 points and group 2 an average of 1.5 points. As mentioned 
above, from a semantical perspective the sequent for group 2 is harder to proof and, therefore, needs a more involved strategy. This is the main reason 
for the only slight improvement in the average grade; a syntactically correct but unsuccessful proof attempt is not graded with more than 2 out of 4
points. Hence, the introduction of the Sequent Calculus Trainer into the process of learning how to correctly apply rules resulted in a 25\%-gain in
points between these two groups. The still low absolute average value achieved by group 2 points out the lack of ``understanding'' of the underlying theory which is not addressed by the use of the Sequent Calculus Trainer. 

\begin{figure}[t]
\begin{center}
\includegraphics[width=\textwidth]{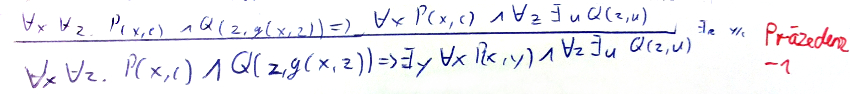}
\end{center}
\caption{Example of the most frequent mistake in group 3.}
\label{fig:solutions_group3}
\end{figure}

Secondly, the results of group 3 seem to worsen in comparison to group 2. Again, the exercise on sequent calculus of group 3 was graded under the same
constraints with 4 points in total. The effort needed to prove this sequent is comparable to the effort for the sequent of group 2, as both need a
similar strategy of replacing the quantified variables in the correct order. However, the outcome is different. In group 3 more students made mistakes in
rule applications although they were also provided with the Sequent Calculus Trainer. They reached an average of 2.7 points. One explanation of
this observation can be found when comparing the mistakes made in rule applications. Out of 19 students who tried to solve the exercise and failed in
rule applications 7 made the same single mistake of misreading the precedence of the conjunction in the succedent, exemplarily shown in
Figure~\ref{fig:solutions_group3}. Such ``near'' solutions were graded with 3 out of 4 points.

\section{Discussion and Future Work}

The improvement in the exam results of the logic course at the University of Kassel in winter term 2013/14 and 2014/15 compared against those achieved in
winter term 2012/13 correlate to the availability and encouragement to actively make use of the Sequent Calculus Trainer. Clearly, the improvement of
results in one particular course can be caused by various reasons; it is well known that the teaching staff has the greatest impact on learning outcomes
(cf. \cite{hattie2008visible}). In our setting the lectures for all groups were given by the same lecturer and the
main part of the exercises was given by the same teaching assistant. The role of the group's composition and the educational background of the students
is of course not to be underestimated. However, the significance of the measured effect was too great to be caused solely by the aforementioned
reasons. This is strongly emphasised by the fact that the differences in the total outcome of the exams are marginal. Nevertheless, it would obviously be
interesting to support this further by a broader empirical evaluation, in particular by considering its effects on students at different universities.

When regarding the effect more deeply, a software for teaching sequent calculus used as an interactive compiler for the language of proof trees perfectly
meets the requirements given in Section~\ref{didactics} for closing the syntactic gap that may be present in students' minds when introducing a new
formal concept. From a different point of view such a piece of software offers a suitable alternative in addressing the rote memory to replace just the
right amount of pen and paper exercises, needed to understand the formalism. In addition, it directly forbids mistakes that would be possible to make
with pen and paper and would have to be manually marked by human teaching assistants.

Clearly, the goal in teaching a calculus for propositional or first-order logic is not just about the simple manipulation of strings. Instead students 
need to learn to fluently understand the properties being expressed by logical formulas, to visualise the classes of structures that are being
represented by them, etc. In other words, students also need to understand the semantics of logical languages and calculi. The Sequent Calculus Trainer
is not meant to address possible deficits in understanding semantics, nor does it do it automatically as the considerations at the end of 
Section~\ref{sec:empirical} show. We do believe, though, that similar improvement in learning success could be achieved for such semantical aspects by
complementing the Sequent Calculus Trainer with a tool that trains the understanding of semantics, for instance using model checking games for
first-order logic \cite{Gradel:2005:FMT:1206819}.

\section*{Acknowledgements}

We would like to thank Angelika Hoffman-Hesse for doing the statistical evaluation of which mistakes were made by how many students as reported above,
and Florian Bruse for acting as a linguistic consultant in the development of the Sequent Calculus Trainer. The work has been financially supported by
the \emph{Service Center Lehre} of the University of Kassel under an \emph{Innovation-in-Teaching} grant.

\bibliographystyle{plain}        
\bibliography{ttl} 
\end{document}